\documentclass[10pt,twocolumn,preprintnumbers,prd,superscriptaddress,nofootinbib,floatfix,showpacs,showkeys,aps]{revtex4-2}

\usepackage{float}
\usepackage{overpic}
\usepackage{graphicx}  
\usepackage{multirow}
\usepackage{natbib}
\usepackage{pifont}
\usepackage{amssymb}
\usepackage{bm}        
\usepackage{amsfonts}  
\usepackage{amsmath}   
\usepackage{amssymb}   
\usepackage[titletoc]{appendix}
\usepackage{color}
\usepackage[english]{babel}
\usepackage{booktabs}
\usepackage{mathtools}
\usepackage{subfigure}
\usepackage{enumerate}
\usepackage{soul}
\definecolor{vividviolet}{rgb}{0.62, 0.0, 1.0}
\definecolor{amaranth}{rgb}{0.9, 0.17, 0.31}
\definecolor{palatinateblue}{rgb}{0.15, 0.23, 0.89}
\definecolor{brightpink}{rgb}{1.0, 0.0, 0.5}
\definecolor{cornflowerblue}{rgb}{0.39, 0.58, 0.93}
\definecolor{deepcarminepink}{rgb}{0.94, 0.19, 0.22}
\definecolor{radicalred}{rgb}{1.0, 0.21, 0.37}

\usepackage{hyperref}
\usepackage{cleveref}
\hypersetup{ linktoc=all,
    colorlinks, linkcolor={palatinateblue},
    citecolor={brightpink}, urlcolor={amaranth}
}

\begin{document}

\title{Gravitational waves from warm inflation in the weak dissipative regime}

\author{Orlando Luongo}
\email{orlando.luongo@unicam.it}
\affiliation{University of Camerino, Via Madonna delle Carceri, Camerino, 62032, Italy}
\affiliation{Istituto Nazionale di Fisica Nucleare (INFN), Sezione di Perugia, Perugia, 06123, Italy}
\affiliation{Department of Nanoscale Science and Engineering, University at Albany SUNY, Albany, NY 12222, USA}
\affiliation{INAF - Osservatorio Astronomico di Brera, Milano, Italy}
\affiliation{Al-Farabi Kazakh National University, Almaty, 050040, Kazakhstan}

\author{Tommaso Mengoni}
\email{tommaso.mengoni@unicam.it}
\affiliation{University of Camerino, Via Madonna delle Carceri, Camerino, 62032, Italy}
\affiliation{Istituto Nazionale di Fisica Nucleare (INFN), Sezione di Perugia, Perugia, 06123, Italy}
\affiliation{INAF - Osservatorio Astronomico di Brera, Milano, Italy}

\author{Paulo M. S\'a}
\email{pmsa@ualg.pt}
\affiliation{Departamento de F\'\i sica, Faculdade de Ci\^encias e Tecnologia, Universidade do Algarve, Campus de Gambelas, 8005-139 Faro, Portugal}
\affiliation{Centro de Investiga\c{c}\~ao e Desenvolvimento em Matem\'atica e Aplica\c{c}\~oes (CIDMA), Universidade de Aveiro, Campus de Santiago, 3810-193 Aveiro, Portugal}

\begin{abstract} 
Previous work on the gravitational-wave background generated in a two-scalar-field cosmological model, in which warm inflation and the dark sector are unified within a single framework, has shown that the resulting spectrum could be potentially detectable by planned next-generation gravitational-wave observatories.
In this work, we extend this analysis to the weak dissipation regime of warm inflation, highlighting how the features of the inflationary scenario play a crucial role in the production of gravitational waves.
The full gravitational-wave energy spectrum is calculated using the formalism of continuous Bogoliubov coefficients.
By comparing our results with those obtained in the strong dissipation regime and with the sensitivity curves of future detectors, we find that the weak dissipation regime improves the prospects for observational detection.
\end{abstract}

\keywords{Gravitational waves. Inflation. Dark matter. Dark energy}

\pacs{04.30.-w, 95.36.+x, 95.35.+d, 98.80.Cq, 03.50.-z}


\maketitle 


\section{Introduction}\label{Section 1}

Direct detection of gravitational waves ten years ago \cite{Abbott:2016} opened new exciting avenues for studying and understanding extreme phenomena in astrophysics and cosmology.
To date, with the available ground-based laser interferometers LIGO (Laser Interferometer Gravitational-Wave Observatory), Virgo, and KAGRA (Kamioka Gravitational Wave Detector), it has been possible to isolate and identify gravitational wave signals from mergers of black holes and neutron stars \cite{KAGRA:2021vkt}.
However, from a theoretical standpoint, it is expected that other astrophysical and cosmological phenomena could produce signals strong enough to be detected with current and upcoming technologies \cite{Bailes:2021tot}.

Of particular significance are primordial gravitational waves, generated in the first moments after the Big Bang \cite{Roshan:2025}, which, when detected, will enable direct observation of a period in the Universe's history that cannot be explored through electromagnetic waves.
Accordingly, the detection of primordial gravitational waves will provide invaluable information about the inflationary period and the subsequent transition to the radiation-dominated era, potentially allowing us to distinguish between competing models and to confirm or refute them.

Recently, we calculated the full gravitational-wave energy spectrum within a two-scalar-field cosmological model that unifies warm inflation, dark matter, and dark energy\footnote{See Refs.~\cite{Odintsov, o1, Liddle:2008, Henriques:2009, Santiago:2011, Lima:2019, Oikonomou:2021, o2, o3, Luongo:2024opv, o4, Paliathanasis:2025mvy} for a broader discussion of unified models. Notably, also unified dark energy-dark matter models, not including the inflationary epoch, have been often proposed in the literature, see e.g. \cite{ude, dunsby:2016, dunsby:2024, sa:2020b}.} \cite{Luongo:2026}.
In our study, we assumed that during the inflationary period, there was significant energy transfer between the scalar fields (one of which acts as the inflaton) and the preexisting radiation bath, therefore considering a strong regime of warm inflation.
The resulting spectrum was then compared with the sensitivity curves of planned next-generation ground- and space-based gravitational-wave observatories.
This article aims to extend this analysis to the case of the weak regime of warm inflation. 
In particular, we investigate how the reduction of dissipative effects during the inflationary stage modifies the production of primordial tensor modes within the same unified cosmological framework.
To this end, we reconsider the dynamical evolution of the two-field system in the regime where the dissipation ratios remain smaller than unity for most of the inflationary epoch and compute the corresponding gravitational-wave energy spectrum by employing the formalism of continuous Bogoliubov coefficients.
This approach enables us to consistently follow graviton production across the entire cosmological history, namely from the onset of inflation up to the present epoch.
By comparing the spectra obtained in the weak regime with those previously derived in the strong dissipative scenario, we assess the impact of the dissipative dynamics on the amplitude and spectral shape of the stochastic gravitational-wave background. 
Further, we contrast the resulting predictions with the projected sensitivities of forthcoming gravitational-wave observatories, \emph{highlighting the frequency intervals in which the signal may become observationally accessible}\footnote{For other studies on the gravitational-wave background and its future detectability, see Refs.~\cite{Romano:2016dpx, Christensen:2018iqi, Cai:2020ovp, Cai:2023ykr}.}.

The paper is organized as follows.
In Sec.~\ref{Section 2}, we briefly describe the two-scalar-field cosmological model and characterize the different dissipative regimes that can occur within it.
In Sec.~\ref{Section 3}, we compute the corresponding full gravitational-wave energy spectra using the continuous Bogoliubov coefficient formalism and compare them with the sensitivity curves of future detectors.
Finally, in Sec.~\ref{Section 4}, we discuss the results and present our conclusions.

\section{The strong and weak dissipation regimes in the two-scalar-field cosmological model}\label{Section 2}

A unified description of inflation, dark matter, and dark energy can be achieved within a two-scalar-field cosmological model \cite{Sa:2020,Sa:2021}, defined by the action
\begin{align}
\label{action}
    S = {} \int& d^4x \sqrt{-g} \bigg[
  \frac{R}{2\kappa^2} - \frac12 (\nabla \phi)^2 \notag \\&- \frac12 e^{-\alpha\kappa\phi} (\nabla \xi)^2
  - e^{-\beta\kappa\phi} V(\xi) \bigg],
\end{align}
where $g$ is the determinant of the metric, $R$ is the Ricci scalar, $\xi$ and $\phi$ are scalar fields, the potential is given by $V(\xi)=V_a+\frac12m^2\xi^2$, $\alpha$ and $\beta$ are dimensionless parameters, and $\kappa\equiv\sqrt{8\pi}/m_p$.

Actions of this type arise naturally in several modified gravity frameworks, such as scalar-tensor theories (e.g. Jordan–Brans–Dicke), $f(R)$ gravity, higher-dimensional theories, and supergravity or string-inspired models. 
In these contexts, conformal transformations between the Jordan and Einstein frames lead to exponential couplings between scalar fields \cite{Berkin:1991nm}, providing a unification framework for the dark sector and inflation.
In our unification scheme, the scalar field $\xi$ is first identified as the inflaton and subsequently as dark matter, while the scalar field $\phi$ plays the role of dark energy.

It is assumed that inflation is of the warm type \cite{Berera:1995, Berera:2023}, implying that the inflaton dissipates its energy into other light degrees of freedom faster than the Hubble expansion rate, allowing the produced particles to thermalize and become radiation \cite{kamali:2023}.
This energy transfer is mediated by a dissipation coefficient $\Gamma$, which, in general, depends on the temperature $T$ of the radiation bath and the inflaton field $\xi$, and is determined by the underlying microscopic model.
Over the years, several warm inflationary models have been proposed \cite{kamali:2023}, in which dissipation coefficients have the form $\Gamma \propto T^a \xi^b$, where $a$ and $b$ are constants.

In our two-scalar-field framework, we adopt a phenomenological and model-independent approach, assuming that: i) both the inflaton $\xi$ and the dark-energy field $\phi$ transfer energy to the radiation bath, ii) the dissipation coefficients $\Gamma_{\xi}$ and $\Gamma_{\phi}$  depend solely on the temperature of the radiation bath, and iii) immediately after the end of the inflationary period, the dissipation coefficients are exponentially suppressed. Under these assumptions, warm inflation can be effectively described by \cite{Sa:2020}
\begin{align}
 \Gamma_{\xi,\phi} = f_{\xi,\phi} \times \left\{
    \begin{aligned}
     & T^p,
     & T\geq T_\texttt{E},
    \\
     & T^p  \exp \left[ 1- \left( \frac{T_\texttt{E}}{T} \right)^q \right],
     & T\leq T_\texttt{E},
    \end{aligned}
 \right.
 \label{gammas}
\end{align}
where $f_\xi$, $f_\phi$, $p$, and $q$ are constants, $T_\texttt{E}$ denotes the temperature of the radiation bath at the end of inflation,
 and $T$ is related to the energy density of radiation by
\begin{equation}
\rho_\texttt{R}=\frac{\pi^2}{30}g_\ast T^4,
\label{rhoT}
\end{equation}
with the effective number of relativistic degrees of freedom chosen to be $g_\ast=106.75$.
Due to this direct coupling between the scalar fields and the radiation bath, the energy density of the latter remains substantial throughout the inflationary period, allowing for a smooth transition to a radiation-dominated era without the need for a distinct post-inflationary reheating phase.

Assuming a flat Friedman-Lema\^{i}tre-Robertson-Walker metric, during the inflationary period and the subsequent transition to a radiation-dominated era (first stage of evolution), the equations of motion for the scalar fields $\xi(t)$ and $\phi(t)$ and for the energy density of radiation $\rho_\texttt{R}(t)$ are given by
\begin{align}
 & \ddot{\xi} + 3 H \dot{\xi} - \alpha\kappa \dot{\phi}\dot{\xi}
 + \frac{\partial V}{\partial \xi} e^{(\alpha-\beta)\kappa\phi}
 = - \Gamma_\xi \dot{\xi} e^{\alpha\kappa\phi}, \label{ddot-xi-1}
  \\
 & \ddot{\phi} + 3 H \dot{\phi} + \frac12 \alpha\kappa
 \dot{\xi}^2 e^{-\alpha\kappa\phi}
 -\beta \kappa V e^{-\beta \kappa \phi}
 = - \Gamma_\phi \dot{\phi}, \label{ddot-phi-1}
  \\
 & \dot{\rho_\texttt{R}} + 4 H \rho_\texttt{R} = \Gamma_\xi
 \dot{\xi}^2 + \Gamma_\phi \dot{\phi}^2, \label{dot-rho-1}
\end{align}
where an overdot denotes a derivative with respect to cosmic time $t$, and the Hubble parameter is
\begin{equation}
    H^2 = \frac{\kappa^2}{3} \bigg(
 \frac{\dot{\phi}^2}{2} + \frac{\dot{\xi}^2}{2} e^{-\alpha\kappa\phi}
 + V e^{-\beta\kappa\phi} + \rho_\texttt{R} \bigg).
 \label{H-1}
\end{equation}

Immediately after the end of the inflationary period, the dissipation coefficients $\Gamma_\xi$ and $\Gamma_\phi$ are exponentially suppressed, becoming negligible.
In the absence of dissipation, radiation decouples from the scalar fields, further evolving as 
\begin{equation}
\rho_\texttt{R}=\rho_\texttt{R0} \left( \frac{a_0}{a} \right)^4,
\end{equation}
where $a$ is the scale factor and the subscript $0$ indicates present-time values.
During this second stage of evolution, the scalar field $\xi$ begins to oscillate rapidly around the minimum of its quadratic potential, behaving like a nonrelativistic pressureless dark matter fluid \cite{Turner:1983} with energy density \cite{Sa:2020}
\begin{equation}
\rho_\xi=\rho_\texttt{DM0} \left(  \frac{a_0}{a} \right)^3 e^{\frac{(\alpha-\beta)\kappa}{2}(\phi-\phi_0)},
    \label{xi 2}
\end{equation}
where $\rho_\texttt{DM0}$ denotes the present-time energy density of dark matter.
Introducing ordinary baryonic matter, described as a perfect fluid with pressure $p_\texttt{BM}=0$ and energy density $
\rho_\texttt{BM} = \rho_\texttt{BM0} (a_0/a)^3$, the equation for the dark energy field becomes
\begin{align}
&\ddot{\phi}+3H\dot\phi-\beta\kappa V_a e^{-\beta\kappa\phi} \notag \\&+ \frac12 \kappa (\alpha-\beta) \rho_\texttt{DM0} \left( \frac{a_0}{a} \right)^3 e^{\frac{(\alpha-\beta)\kappa}{2}(\phi-\phi_0)} = 0\,,
    \label{ddot-phi-2}
\end{align}
where the Hubble parameter is now given by
\begin{align}
  H^2=&\frac{\kappa^2}{3} \bigg[ \frac{\dot{\phi}^2}{2} + V_a e^{-\beta\kappa\phi} + \rho_\texttt{R0} \left( \frac{a_0}{a} \right)^4 \notag\\&+ \left( \rho_\texttt{BM0} +\rho_\texttt{DM0}e^{\frac{(\alpha-\beta)\kappa}{2}(\phi-\phi_0)} \right) \left( \frac{a_0}{a} \right)^3\bigg].
    \label{H-2}
\end{align}

Agreement with current cosmological measurements \cite{planc} requires  $\rho_\texttt{BM0}=8.19\times10^{-125}\,m_p^4$, $\rho_\texttt{DM0}=4.25\times10^{-124}\,m_P^4$,
$\rho_\texttt{DE0}\equiv\frac12\dot{\phi}_0^2+V_a e^{-\beta\kappa\phi_0}=1.13\times10^{-123}\,m_p^4$, and $\rho_\texttt{R0}=9.02\times 10^{-128}\,m_p^4$, corresponding to a Hubble constant $H_0=67\,\mbox{km}\,\mbox{s}^{-1}\mbox{Mpc}^{-1}$.

After a radiation-dominated era encompassing primordial nucleosynthesis, cold dark matter, together with ordinary baryonic matter, begins to dominate the dynamics of the Universe, giving rise to a matter-dominated era, long enough for structure formation to occur.

Finally, in recent times, the dark energy field $\phi$ emerges as the dominant component of the Universe, giving rise to a second era of accelerated expansion.

The intensity of the energy transfer from the scalar fields $\xi$ and $\phi$ to the radiation bath determines the warm inflationary regime, which is characterized by the dissipation ratios
\begin{equation}\label{eq dissipation ratios}
   Q_\xi=\frac{\Gamma_\xi}{3H} e^{\alpha\kappa\phi} \quad \mbox{and} \quad Q_\phi=\frac{\Gamma_\phi}{3H}.
\end{equation}
If $Q_\xi,Q_\phi>1$, the regime is dubbed strong; if $Q_\xi,Q_\phi<1$, the regime is weak (discussions on the dissipative regimes can be found in Refs.~\cite{Gupta:2003au, Gupta:2005nh, Trivedi:2020ljd, Cheng:2024uvn, Ito:2025lcg, Yeasmin:2025ael}). We can also have $Q_\xi>1$, $Q_\phi<1$  and $Q_\xi<1$, $Q_\phi>1$, which we refer to as the strong-weak and the weak-strong regimes, respectively.

From the inflaton dynamics,  Eq.~\eqref{ddot-xi-1}, the effective friction term reads $3H[1-\alpha\kappa\dot{\phi}/(3H)+Q_\xi]$, implying that cosmic expansion dominates or is subdominant to the additional contributions when
\begin{equation}
      Q_\xi- \frac{\alpha\kappa \dot{\phi}}{3H}  <1\quad \mbox{or} \quad
      Q_\xi- \frac{\alpha\kappa \dot{\phi}}{3H} >1,
\end{equation}
respectively.
Since $\alpha\kappa\dot{\phi}/(3H) \ll Q_\xi$, as revealed by our numerical simulations (see Sect.~\ref{Section 3}), the second term on the left-hand side of the above inequalities can be safely neglected in the following. 

Notably, the additional exponential term in the dissipation ratio $Q_\xi$ generalizes the standard warm inflation scenario and originates from the dissipation term on the right-hand side of Eq.~\eqref{ddot-xi-1} within our multi-field framework.
In Refs.~\cite{Sa:2020, Luongo:2026}, this term was omitted, as the adopted parameter values and initial conditions make it almost equal to unity throughout the inflationary epoch. 
Here, however, we keep it to emphasize the role of the non-minimal coupling between the scalar fields $\xi$ and $\phi$ in modulating the warm inflationary regime.

In a previous work \cite{Luongo:2026}, we calculated the full gravitational-wave energy spectrum arising in the two-scalar-field cosmological model described above, choosing all our scenarios to correspond to the strong regime.
Here, we will extend this analysis to the weak regime, considering scenarios where one or both dissipation ratios are less than one throughout most of the inflationary period.

During warm inflation, the energy density of radiation, $\rho_\texttt{R}$, is subdominant, but can still satisfy the condition $\rho_\texttt{R}^{1/4}>H$, which, due to thermalization, translates into $T>H$. Using Eqs.~(\ref{rhoT}) and (\ref{eq dissipation ratios}), the conditions for the occurrence of a weak regime, $Q_\xi<1$ and $Q_\phi<1$, becomes  
\begin{align}
\frac{3}{f_\xi } \left( \frac{\pi^2 g_\ast}{30\rho_\texttt{R}} \right)^{\frac{p-1}{4}} e^{-\alpha\kappa\phi} >\frac{T}{H}>1,
\\
\frac{3}{f_\phi } \left( \frac{\pi^2 g_\ast}{30\rho_\texttt{R}} \right)^{\frac{p-1}{4}} >\frac{T}{H}>1,
\end{align}
revealing that a transition from a strong to a weak dissipative regime can be achieved by varying the parameters $f_\xi$, $f_\phi$, and $p$ ($\alpha$ is constrained by observational data). In our numerical simulations (see Sect.~\ref{Section 3}), we have chosen to vary $f_\xi$ and $f_\phi$, although varying $p$ instead would lead to an equivalent scenario.
Note that the parameter $q$, introduced in Eq.~(\ref{gammas}), contrary to $p$, does not influence the nature of the dissipation regime; it only controls how rapidly the dissipation coefficients are suppressed at the transition from inflation to the radiation-dominated era.
Given that the production of gravitational waves during this period is incomparably lower than during inflation \cite{Luongo:2026}, the choice of parameter $q$ has a negligible impact on the amplitude of the gravitational wave energy spectrum; therefore, this parameter can be fixed.

In what follows, for the convenience of the numerical calculations, instead of the cosmic time $t$, we will use a new ``time" variable, $u=-\ln(a_0/a)$, related to the number of e-folds of expansion.

In this way, all the history of the Universe, from the beginning of the inflationary period to the present time, can be compressed between $u_i\approx-135$ and $u_0=0$ (under the assumption that we have $70$ e-folds of expansion during inflation).

In this new variable, Eqs.~\eqref{ddot-xi-1}-\eqref{H-1} for the first stage of evolution (inflationary period and transition to a radiation-dominated era) become \cite{Sa:2020, Luongo:2026}
\begin{widetext}
\begin{align}
    \hspace{-0.5mm}  \xi_{uu} &= - \bigg\{
     \bigg[ \frac{\ddot{a}}{a} + 2 \bigg( \frac{\dot{a}}{a} \bigg)^2
           +\frac{\dot{a}}{a} \Gamma_\xi e^{\alpha\kappa \phi} \bigg]\xi_u
    - \alpha\kappa  \bigg( \frac{\dot{a}}{a} \bigg)^2 \phi_u \xi_u
    + m^2 \xi e^{(\alpha-\beta)\kappa \phi}
  \bigg\} \bigg( \frac{\dot{a}}{a} \bigg)^{-2},
      \label{Eq-xi-s1}
  \\
   \hspace{-1.0mm}\phi_{uu} &=  - \bigg\{
     \bigg[ \frac{\ddot{a}}{a} + 2 \bigg( \frac{\dot{a}}{a} \bigg)^2
           +\frac{\dot{a}}{a} \Gamma_\phi
     \bigg]\phi_u
  + \frac{\alpha\kappa}{2} \bigg( \frac{\dot{a}}{a} \bigg)^2 \xi_u^2
  e^{-\alpha\kappa\phi}
   -\beta\kappa \left( V_a +\frac12 m^2 \xi^2 \right)
   e^{-\beta\kappa\phi}
  \bigg\} \bigg( \frac{\dot{a}}{a} \bigg)^{-2},
      \label{Eq-phi-S1}
  \\
   \hspace{0mm} \rho_{\texttt{R}u} &= - 4 \rho_\texttt{R}
  + \frac{\dot{a}}{a} \left( \Gamma_\xi \xi_u^2 + \Gamma_\phi \phi_u^2 \right),
      \label{Eq-rhoR-S1}
\\
   \left( \frac{\dot{a}}{a} \right)^2 &=2\kappa^2
 \frac{ \left( V_a + \frac12 m^2 \xi^2 \right) e^{-\beta\kappa\phi} +
 \rho_\texttt{R} }{6 - \kappa^2 \phi_u^2
 - \kappa^2 \xi_u^2 e^{-\alpha\kappa\phi}},
 \label{Eq-dota-s1}
\\
 \frac{\ddot{a}}{a} &= \frac{\kappa^2}{3}
 \Bigg\{ \frac{2\kappa^2 \left[ \left( V_a+\frac12 m^2 \xi^2 \right)
 e^{-\beta\kappa\phi} + \rho_\texttt{R} \right](\phi_u^2 + \xi_u^2 e^{-\alpha\kappa\phi})}{\kappa^2 \phi_u^2 + \kappa^2 \xi_u^2 e^{-\alpha\kappa\phi} - 6} + \left( V_a + \frac12 m^2 \xi^2 \right)
  e^{-\beta\kappa\phi} - \rho_\texttt{R}
    \Bigg\},
   \label{Eq-dotdota-s1}
\end{align}
while Eqs.~\eqref{ddot-phi-2} and \eqref{H-2} for the second stage of evolution (radiation-, matter-, and dark energy-dominated eras) are given by
  \begin{align}
  \phi_{uu} &=  - \bigg\{
  \bigg[ \frac{\ddot{a}}{a} + 2 \bigg( \frac{\dot{a}}{a} \bigg)^2 \bigg] \phi_u
  -\beta\kappa V_a e^{-\beta\kappa\phi} + \rho_\texttt{DM0}\frac{(\alpha-\beta)\kappa}{2}
  e^{\frac{(\alpha-\beta)\kappa}{2}(\phi-\phi_0)} e^{-3u} \bigg\} \left(
\frac{\dot{a}}{a} \right)^{-2},\label{Eq-phi-s2}
\\
\left( \frac{\dot{a}}{a} \right)^2 &= {}
 2\kappa^2 \bigg[ V_a e^{-\beta\kappa\phi}
 + \left( \rho_{\texttt{BM}0}
 + \rho_\texttt{DM0} e^{\frac{(\alpha-\beta)\kappa}{2}(\phi-\phi_0)} \right) e^{-3u} + \rho_{\texttt{R}0}e^{-4u} \bigg]
 \left( 6-\kappa^2 \phi_u^2 \right)^{-1},
\label{Eq-dota-s2}
\\
 \frac{\ddot{a}}{a} &= {} \frac{\kappa^2}{6} \bigg\{
 4\kappa^2 \bigg[ V_a e^{-\beta\kappa\phi}
     + \left( \rho_{\texttt{BM}0}
     + \rho_\texttt{DM0} e^{\frac{(\alpha-\beta)\kappa}{2}(\phi-\phi_0)} \right) e^{-3u} + \rho_{\texttt{R}0} e^{-4u} \bigg]
 \phi_{u}^2 \left( \kappa^2 \phi_u^2 - 6 \right)^{-1}
 + 2 V_a e^{-\beta\kappa\phi}\notag\\&\qquad  - \left( \rho_{\texttt{BM}0}
   + \rho_\texttt{DM0} e^{\frac{(\alpha-\beta)\kappa}{2}(\phi-\phi_0)} \right) e^{-3u}
   - 2\rho_{\texttt{R}0} e^{-4u} \bigg\},
 \label{Eq-dotdota-s2}
    \end{align}
\end{widetext}
where the subscript $u$ denotes a derivative with respect to this variable.
Equations~\eqref{Eq-dotdota-s1} and \eqref{Eq-dotdota-s2} have been included because it is computationally convenient to consider $\dot a/{a}$ and ${\ddot a}/{a}$ as independent functions of the new variable $u$.

\section{Gravitational-wave energy spectra} \label{Section 3}

To calculate the full gravitational-wave energy spectrum within the two-scalar-field cosmological model described in the previous section, we resort to the formalism of continuous Bogoliubov coefficients\footnote{This formulation is derived from gravitational particle production~\cite{Parker:69}, which has numerous cosmological applications; see, for instance, Refs.~\cite{Ford:2021syk, Belfiglio:2024swy}.}.

According to this formalism, gravitational waves are interpreted as gravitons, whose creation and annihilation operators evolve through continuous Bogoliubov transformations, characterized by the Bogoliubov coefficients $\alpha_k$ and $\beta_k$ (see Refs.~ \cite{Moorhouse:94, Mendes:95, Henriques:2007, Sa:2008, Sa:2009, Sa:2010, Sa:2012, Bouhmadi-Lopez:2010, Bouhmadi-Lopez:2013, Morais:2014, Luongo:2026} for details of this formalism and its application to cosmology).

The gravitational-wave energy density parameter as a function of the present-day frequency $f_0$ is given by 
\begin{equation}
\Omega_{\texttt{GW}}(f_0)=\frac{128\pi^3\hbar G}{3 c^5 H^2_0} f_0^4 \,(|\beta_{k}|^2)_0,
\end{equation}
where $\hbar$ is the reduced Planck constant, $G$ is the Newton constant, $c$ is the speed of light, and $(|\beta_k|^2)_0=(X_0-Y_0)(X_0^\ast-Y_0^\ast)/4$ is the number of created gravitons evaluated at the present time, $u_0=0$. The continuous functions $X(u)$ and $Y(u)$ are determined by the set of differential equations,
\begin{align}
    &X_u=-2\pi if_0 e^{-u} \frac{Y}{\dot{a}/a}, \label{SODE-XYua}
\\
    &Y_u=-\frac{i}{2\pi f_0} e^u \left[ 4\pi^2f_0^2 e^{-2u} -\frac{\ddot{a}}{a}-\left( \frac{\dot{a}}{a} \right)^2 \right] \frac{X}{\dot{a}/a}, \label{SODE-XYub}
\end{align}
where $\dot{a}/a$ and $\ddot{a}/a$ are determined from Eqs.~\eqref{Eq-xi-s1} - \eqref{Eq-dotdota-s1} and \eqref{Eq-phi-s2} - \eqref{Eq-dotdota-s2} for the first and second stages of evolution, respectively. 

Let us now specify the parameter values and initial conditions for four scenarios corresponding to the strong, weak, strong-weak, and weak-strong regimes.

The dimensionless parameters $\alpha$ and $\beta$, introduced in action \eqref{action}, can be constrained by observational data.
Using a Monte Carlo Markov chain analysis against low-redshift datasets, it was established \cite{Luongo:2022} that
\begin{equation}
    \alpha = 0.36^{+0.18}_{-0.26} \quad \mbox{and} \quad \beta= 0.01^{+0.34}_{-0.24} \,.
    \label{fit params}
\end{equation}
We adopt the mean values, namely, $\alpha = 0.36$ and $\beta= 0.01$.
For the bare mass of the scalar field $\xi$, we take $m=10^{-5}\,m_p$. 
For the parameters determining the temperature dependence of the dissipation coefficients, we choose $p=1$ and $q=2$.
In all scenarios, the initial conditions are chosen to be $\phi_i=10^{-3}\,m_p$,
$\phi_{u,i}=10^{-5}\,m_p$,
$\xi_{u,i}=10^{-2}\,m_p$, and $\rho_{\texttt{R}, i}=0.25\times10^{-12}\,m_p^4$.

It is worth emphasizing that the constraints on $\alpha$ and $\beta$ are derived from observational data describing the late-time evolution of the Universe.
In both the strong and weak dissipative regimes of warm inflation, at the end of the first stage of evolution, when the energy transfer ceases, the resulting energy densities of the scalar fields are compatible with current observational bounds on dark energy and dark matter. Therefore, the above values  for $\alpha$ and $\beta$ can be consistently adopted in both scenarios.

As discussed in Sect.~\ref{Section 2}, the nature of the dissipative regime is determined by the choice of the parameters $f_\xi$, $f_\phi$, and $p$.
To transition from a strong regime to a weak one, we choose to fix $p$ and vary $f_\xi$ and $f_\phi$, whose values are presented in Table ~\ref{table-1}.

To ensure $70$ e-folds of expansion during inflation, the initial value of the $\xi$ field must be varied from case to case; these values are also shown in Table ~\ref{table-1}.

Finally, to ensure that at present time the energy density of the field $\phi$ is equal to the observed value for dark energy, $\rho_{\phi0}=\rho_\texttt{DE0}$, the constant $V_a$ must be slightly adjusted for each of the scenarios under consideration; its values are also shown in Table ~\ref{table-1}.

\begin{table}[t]
\centering
\renewcommand{\arraystretch}{1.2}
\begin{tabular}{c||c|c|c|c}
\hline\hline
Scenario & $f_\xi$ & $f_\phi$ & $\xi_i\,[m_p]$ &  $V_a\,[10^{-123}\,m_p^4]$  \\
\hline\hline
\emph{strong} & 2.0 & 2.0 & 0.745 & 1.098  \\
\hline
\emph{weak} & 0.1 & 0.1 & 2.924 & 1.102  \\
\hline
\emph{strong-weak} & 2.0 & 0.1 & 0.746 & 1.101  \\
\hline
\emph{weak-strong} & 0.1 & 2.0 & 2.861 & 1.099\\
\hline\hline
\end{tabular}
\caption{Model parameters and initial conditions for the four scenarios under consideration.}
\label{table-1}
\end{table}

\begin{figure*}[ht]
    \centering
    \begin{overpic}[scale=1.3,grid=false]{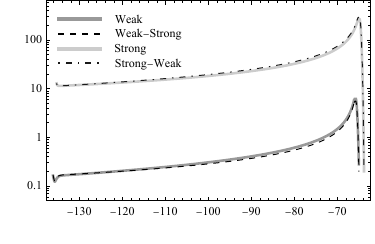}
\put(56,2){\makebox(0,0){$u$}}
\put(2,31){\rotatebox{90}{$\log({ Q}_\xi)$}}
\end{overpic}
    \begin{overpic}[scale=1.3,grid=false]{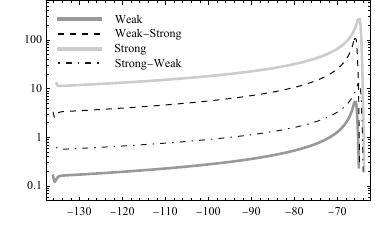}
\put(56,2){\makebox(0,0){$u$}}
\put(2,31){\rotatebox{90}{$\log({ Q}_\phi)$}}
\end{overpic}
    \caption{ Evolution of the dissipation ratios $Q_\xi$ and $Q_\phi$, for the four scenarios under consideration. At the end of the inflationary period ($u\approx-65$), the dissipation coefficients $\Gamma_\xi$ and $\Gamma_\phi$ are exponentially suppressed, implying that soon afterward the dissipation ratios become negligible. This marks the end of the first stage of evolution.
    }
    \label{fig dissipation ratios}
\end{figure*}

Within these parameter choices, the evolution of the dissipation ratios, defined in Eq.~\eqref{eq dissipation ratios}, is shown in Fig.~\ref{fig dissipation ratios}.
In particular, the four scenarios are chosen such that the inflaton $\xi$ exhibits two well-defined regimes, strong and weak.
Since the field $\xi$ dominates the inflationary dynamics and hence the evolution of the Hubble parameter, the dissipation ratio of the field $\phi$ is affected by the inflaton, leading to four different configurations: weak, strong-weak, weak-strong, and strong.
However, as we will discuss later, the behavior of the field $\phi$ has no noticeable impact on the production of gravitational waves, which is determined by the dissipation regime of the inflaton field $\xi$.

\begin{figure*}[ht]
\centering
    \begin{overpic}[scale=1.25,grid=false]{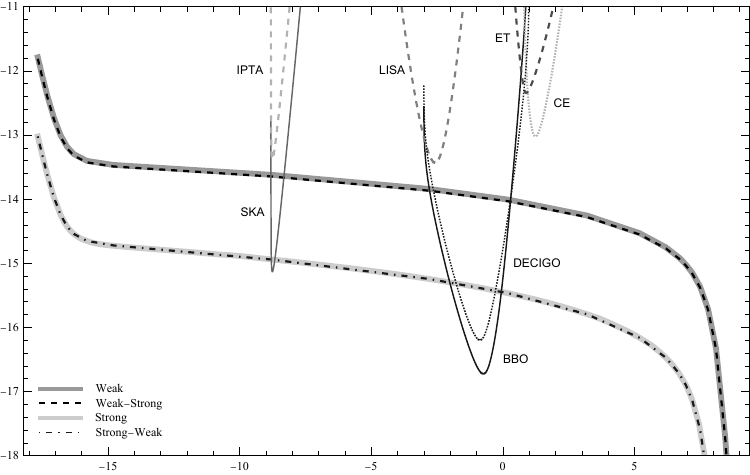}
\put(53.5,-1.5){\makebox(0,0){$\log(f/{\rm Hz})$}}
\put(-3,30.5){\rotatebox{90}{$\log({ \Omega}_{GW})$}}
\end{overpic}
\caption{Full gravitational-wave energy spectra for the four scenarios under consideration. The amplitude of $\Omega_{\texttt{GW}}$ increases by over an order of magnitude across the entire allowed frequency range in the weak scenario as compared to the strong one. The spectra for the strong-weak and weak-strong scenarios almost exactly coincide with the spectra for the strong and weak cases, respectively. The obtained spectra are superimposed on the sensitivity curves \cite{Schmitz:2020syl} (for the repository with the sensitivity curves, see Ref.~\cite{repository}) of LISA, BBO, CE, ET, SKA, IPTA, and DECIGO. In the frequency bands $(10^{-9}-10^{-8})\,\mbox{Hz}$ and $(10^{-3}-1)\,\mbox{Hz}$, the gravitational waves signals are potentially detectable.}
\label{Fig spectra}
\end{figure*}

By considering frequencies corresponding to wavelengths ranging from the present Hubble horizon ($f_0\approx10^{-16.9}\,{\rm Hz}$) up to the horizon scale at the end of inflation ($f_0\approx10^{8.4}\,{\rm Hz}$), we compute the resulting gravitational-wave energy spectra for our four scenarios\footnote{The spectrum corresponding to the strong regime is the one obtained in Ref.~\cite{Luongo:2026}.}, which are shown in Fig.~\ref{Fig spectra}, superimposed on the sensitivity curves of some planned next-generation ground- and space-based gravitational wave observatories ~\cite{Moore:2014lga, Ringwald:2020vei}, namely, Laser Interferometer Space Antenna (LISA), Big Bang Observer (BBO), Cosmic Explorer (CE), Einstein Telescope (ET), Square Kilometer Array (SKA), International Pulsar Timing Array (IPTA), and Deci-Hertz Interferometer Gravitational Wave Observatory (DECIGO).

An analysis of these four spectra reveals that in the weak scenario, the amplitude of the gravitational-wave energy density parameter $\Omega_{\texttt{GW}}$ increases by more than an order of magnitude across the entire range of allowed frequencies, making it easier to detect such waves.
This larger amplitude in the weak regime can be understood as follows. Gravitational waves are mainly generated from the inflationary background. However, in warm inflation, the dissipative dynamics continuously transfer energy from the inflaton sector into radiation, thereby reducing the energy available to excite tensor modes. This effect results to be maximal in the strong dissipative regime.

By contrast, scalar perturbations in standard warm inflation are significantly affected by thermal fluctuations of the radiation bath \cite{Lima:2019, kamali:2023}, which tend to enhance the scalar power spectrum in the strong dissipative regime \cite{Bastero-Gil:2016qru}. 
As a consequence, the tensor-to-scalar ratio $r \equiv P_T/P_S$ may exhibit a nontrivial dependence on the dissipative regime. 
In particular, while tensor modes are suppressed in the strong regime, scalar modes are simultaneously amplified, leading to an additional reduction of $r$. 
This interplay suggests that a combined analysis of scalar and tensor perturbations could provide a powerful probe into different dissipative regimes in multi-field warm inflation.
Although warm inflation is generally consistent with current observational constraints \cite{Planck:2018jri, Visinelli:2016rhn, Benetti:2016jhf, Montefalcone:2022jfw, B:2025koi}, a dedicated analysis of scalar perturbations within our two-field framework is required to get quantitative conclusions, and it will be the subject of future work.

The four spectra shown in Fig.~\ref{Fig spectra} also reveal that, for fixed parameters, the dissipative regime of the dark energy scalar field $\phi$ minimally affects the amplitude of $\Omega_{\texttt{GW}}$. Instead, this amplitude is almost entirely determined by the dynamics of the inflaton $\xi$, particularly by its dissipative regime.

Our numerical simulations indicate that when we vary the values of the parameters $\alpha$ and $\beta$ within the ranges permitted by Eq.~\eqref{fit params} in the weak regime scenario, this regime gradually changes to a strong one for both fields $\xi$ and $\phi$. This implies a decrease in the amplitude $\Omega_{\texttt{GW}}$, that is, the corresponding spectra approach the spectrum obtained in the strong regime.

\section{Conclusions}\label{Section 4}

In this work, we extended the recent analysis of Ref.~\cite{Luongo:2026}, where gravitational waves generated within a two-scalar-field cosmological model were investigated.
In that study, several parameter choices were explored, albeit the analysis was restricted to the strong dissipation regime.

Here, we considered the same theoretical framework but focused instead on the weak dissipation regime.
In particular, we considered a weak ($Q_\xi,Q_\phi<1$), strong-weak ($Q_\xi>1$, $Q_\phi<1$) and weak-strong ($Q_\xi<1$, $Q_\phi>1$) regimes, along with the strong ($Q_\xi,Q_\phi>1$) analyzed in Ref.~\cite{Luongo:2026}.

By adopting the formalism of continuous Bogoliubov coefficients, we computed the full gravitational-wave energy spectrum, spanning frequencies from the minimum value corresponding to the present Hubble horizon up to the maximum frequency set by the end of the inflationary phase.

Our analysis indicates that the strength of the dissipative regime of the dark energy field $\phi$ has a negligible impact on the amplitude of the gravitational-wave energy density parameter $\Omega_{\texttt{GW}}$. Therefore, the gravitational-wave spectrum is determined almost entirely by the dynamics of the inflaton $\xi$, particularly its dissipative regime.

In the weak regime scenario, the amplitude of $\Omega_{\texttt{GW}}$ increases by more than an order of magnitude across the entire allowed frequency range, when compared to the strong regime scenario.
This circumstance enhances the prospects for future detection of these primordial gravitational waves and may even allow for the discrimination between alternative models of warm inflation with different dissipation regimes.

Our work describes in detail tensor perturbations within a two-scalar-field cosmological model that unifies inflation, dark energy, and dark matter. However, a comprehensive analysis, allowing us to extract additional information and provide further observational tests for the model, also requires the calculation of scalar perturbations. Since the consideration of a warm inflationary model with multiple scalar fields requires a non-trivial generalization of scalar perturbation theory, this analysis is left for future work.

In conclusion, the development of gravitational-wave cosmology, in combination with scalar-perturbation analyses, will provide a crucial tool to probe the very early Universe. From this perspective, further studies of primordial gravitational waves are required in order to obtain a consistent and comprehensive picture.

\section*{Acknowledgments}

TM acknowledges hospitality to the University of Algarve during the time in which this article was conceived and finalized.
PS acknowledges support from CIDMA-Center for Research and Development in Mathematics and Applications under FCT grants UID/04106/2025 (\url{https://doi.org/10.54499/UID/04106/2025}) and UID/PRR/04106/2025 (\url{https://doi.org/10.54499/UID/PRR/04106/2025})

\end{document}